\documentclass[conference]{IEEEtran}

\usepackage{cite}
\usepackage{amsmath,amssymb,amsfonts}
\usepackage{algorithmic}
\usepackage{graphicx}
\usepackage{textcomp}
\usepackage[numbers]{natbib}

\usepackage[hyphens,spaces,obeyspaces]{url}

\usepackage{amsmath}

\usepackage{tikz}

\usetikzlibrary{positioning}

\tikzset{basic/.style={draw,fill=blue!10,text width=1em,text badly centered}}
\tikzset{input/.style={basic,circle}}
\tikzset{weights/.style={basic,rectangle}}
\tikzset{functions/.style={basic,circle,fill=blue!10}}

\usetikzlibrary{shapes,arrows}
\tikzstyle{block} = [draw, fill=blue!20, rectangle, minimum height=3em, minimum width=6em]
\tikzstyle{sum} = [draw, fill=blue!20, circle, node distance=1cm]
\tikzstyle{input} = [coordinate]
\tikzstyle{output} = [coordinate]
\tikzstyle{pinstyle} = [pin edge={to-,thin,black}]

\newcommand\Tstrut{\rule{0pt}{2.6ex}}         
\newcommand\Bstrut{\rule[-0.9ex]{0pt}{0pt}}   

\def\BibTeX{{\rm B\kern-.05em{\sc i\kern-.025em b}\kern-.08em
   T\kern-.1667em\lower.7ex\hbox{E}\kern-.125emX}}

\makeatletter

\def\ps@IEEEtitlepagestyle{%
   \def\@evenfoot{}%
}

\title{MLP$_{df}$: An Effective Machine Learning Based Approach for PDF Malware Detection}

\begin{document}

\author{\IEEEauthorblockN{\textbf{Jason Zhang}\textit{, Ph.D.}}
\IEEEauthorblockA{\textit{Senior Threat Researcher}}
\textit{Sophos, Abingdon OX14 3YP, U.K.}\\
jason.zhang@sophos.com
}

\thispagestyle{plain}
\pagestyle{plain}

\maketitle

\begin{abstract}
Due to the popularity of portable document format (PDF) and increasing number of vulnerabilities in major PDF viewer applications, malware writers continue to use it to deliver malware via web downloads, email attachments and other methods in both targeted and non-targeted attacks. The topic on how to effectively block malicious PDF documents has received huge research interests in both cyber security industry and academia with no sign of slowing down. In this paper, we propose a novel approach based on a multilayer perceptron (MLP) neural network model, termed MLP$_{df}$, for the detection of PDF based malware. More specifically, the MLP$_{df}$ model uses a backpropagation algorithm with stochastic gradient decent search for model update. A group of high quality features are extracted from two real-world datasets which comprise around $105000$ benign and malicious PDF documents. Evaluation results indicate that the proposed MLP$_{df}$ approach exhibits excellent performance which significantly outperforms all evaluated eight well known commercial anti-virus scanners with a much higher true positive rate of $95.12\%$ achieved while maintaining a very low false positive rate of $0.08\%$.
\end{abstract}


\section{Introduction}
\subsection{Why Targeting PDF?}
Portable document format (PDF) is one of the most commonly used file formats for electronic documents exchange across applications and platforms. The detailed PDF specification can be found in \cite{Adobe}. A few factors help PDF become one of malware writers' favourite file formats to spread malicious content: i) it is widely used by general users in both work and non-work environments. Typical examples include academic articles, technical reports, design documents and electronic receipts; ii) it is independent of operating systems (OS) and platforms. One can open a PDF file with a standalone PDF viewer or within a modern web browser (with PDF viewer plug-in) on a Windows machine, a Linux system or a mobile device; iii) it is an extremely flexible file format. PDF supports various types of data in addition to text, e.g. JavaScript, Flash, media files, interactive forms, or linking to external files and uniform resource locators (URLs), etc. Furthermore, various encoding and compression methods can be used for the purpose of reducing file size, hiding sensitive content, or both; and iv) it is stealth and elegant. Normally PDF files are believed to be less suspicious than executable files. It is a common security practice for an IT administrator to define a policy to block executable files from staff email attachments or web downloads, but it is rare to block PDF documents in such a manner.

Because of the huge popularity and flexibility of PDF file format, it also opens up many ways for attackers to propagate malware via PDF documents. 
\subsection{PDF Based Malware}
PDF based attacks typically fall into two categories - phish and exploits. Phishing attacks are commonly seen in emails. A typical example can be a PDF based order confirmation or delivery receipt attached to an email claiming to be sent from a well known online shopping portal or logistic company. The text content in such emails provides little value except using social engineering to entice email receivers to open attached phishing PDF files. Such PDF files are typically single page long with some social engineering texts and a phishing URL which leads to a suspicious website for the purpose of personal information and account credential harvest, or malicious file download, etc. As compared to plain text based phishing attacks, it is more challenging to detect PDF based phish as PDF documents are binary or mixture of binary and ASCII texts. This is one of the reasons why PDF based phishing attacks are increasingly popular. The motivation behind it is nothing different from typical phishing attacks. The harvested information from victims could be either used by the attackers themselves or sold on black markets for business potential of the so-called shadow economy.

Another common way to spread malware via PDF is to exploit a vulnerable PDF viewer application in order to execute malware payload which can be either embedded in a PDF file or downloaded. In most cases, this is achieved by taking advantage of JavaScript supported by PDF to trigger specific vulnerabilities and then execute code of an attacker's choice. Typically this involves obfuscation and memory manipulation techniques like buffer overflow, return oriented programming (ROP) and heap spraying via encoded shellcode \cite{ZhangR11b}. Such malicious content can be encapsulated in a single object,  scattered across multiple objects in a PDF document, or in the form of lengthy encoded strings (like hex or decimal encoding) stored in PDF \emph{Info} dictionary metadata tags (e.g. Title, Subject, Author). To maximize an attack success rate, a malicious PDF document could be used to target multiple vulnerabilities. For example, an attack detected by Sophos targets four vulnerabilities of a popular PDF reading application \cite{Z15}: \textit{Util.printf}() (CVE-2008-2992), \textit{Collab.getIcon} (CVE-2009-0927), \textit{Collab.collectEmailInfo} (CVE-2007-5659) and \textit{Escript.api plugin media player} (CVE-2010-4091). Each of them affects a certain version of the vulnerable application depending on the version installed on a victim's machine. The detailed information for these vulnerabilities will not be discussed here. It's readily available online. Apart from the vulnerability exploits discussed above, one can also use other PDF features like \emph{/OpenAction}, \emph{/AA} and \emph{/launch} to automatically launch a malicious application or run different commands depending on an OS. 

\subsection{Traditional Detection}
There are many approaches developed to block PDF based attacks, varying from static detection like signature match to dynamical analysis using sandbox technologies. One of the advantages of signature based detection is that it is good at detecting known malware with relatively low FP rates. On the other hand, as signature based detection normally uses byte sequences to match specific malware, it is not robust when dealing with zero-day attacks or malware variants. This puts great challenges to AV scanners relying heavily on signature detection. An alternative approach is based on dynamical analysis, or termed behavior-based malware detection. It uses sandbox technology to add an extra layer of detection. Rather than relying on byte sequences match, it monitors behaviour exhibited from a PDF file when opening it in a controlled environment and a detection decision will be made if certain behaviour is observed. This greatly improves the detection rate for attacks even with highly obfuscated content like JavaScript. It is worth noting that sandbox based technology only works well if an observed file performs actual malicious operations as if running in a real environment. As mentioned above, malware authors already found many ways to bypass signature based detection with techniques like obfuscation, encryption, etc. Similarly, they constantly innovate ways to evade sandbox tools with so called anti-sandbox techniques. For example, if the presence of a sandbox environment is detected, it only exhibits benign behavior or switches to a \emph{sleep} mode. There also exist other limitations. Some sandbox tools only deal with specific types of PDF attacks like MDScan for JavaScript \cite{TzermiasSPM11C}, Nozzel for heap spraying \cite{RatanaworabhanLZ09C}, or it only records dynamic behavior of a system and still requires manual analysis to form a detection decision as in the case of CWSandbox \cite{WillemsHF07}.

\subsection{ML Based Detection}\label{MLD}
A machine learning algorithm is an algorithm that is able to learn from data \cite{GoodfellowBC16}. A more precise definition from Tom Mitchell says ``computer program is said to learn from experience E with respect to some class of tasks T and performance measure P, if its performance at tasks in T, as measured by P, improves with experience E" \cite{Mitchell97}. There is a very wide variety of experiences E, tasks T, and performance measures P depending on applications. In the case of PDF attacks detection, the task T herein is to classify PDF documents as malicious or benign. The experience E is a collection of pre-classified PDF documents given to the algorithm to learn. The measured performance P will hopefully improve when learning from updated experience E (new PDF documents).

In the past few years, ML, deep learning (DL) and artificial intelligence (AI) have been a hot subject of research and application across industries due to modern hardware with increased computational power, available big datasets, and improved algorithms, etc. This has led to many breakthroughs from image classification, speech recognition to autonomous driving. Similar trend is observed in cyber security and major security vendors such as Sophos have already introduced ML based detection together with traditional approaches to build a multi-layered arsenal of protection in modern cyber security \cite{Sophos17}. 

Recent work has also witnessed several ML applications in PDF based malware detection. Given the fact that vast majority of PDF based exploits are JavaScript related, many of the ML algorithms are designed to detect malicious JavaScript code in PDF files. An example of these is Wepawet which comprises the approach from CWSandbox and a classification system \cite{Wepawet}. The tool is mainly used to detect malicious URLs and JavaScript based PDF attacks. Laskov and Srndic \cite{LaskovS11C} proposed an approach called PJScan which uses static analysis and support vector machines (SVM) to classify JavaScript inside PDF documents with modest detection rates achieved. 

There exist a few learning-based approaches for general PDF malware detection instead of JavaScript focused, with pros and cons. In Munson and Cross's report \cite{CrossM11T}, a decision tree based ensemble learning algorithm is presented to classify PDF files. An instrumented PDF reading application is used to extract features as input to the learning model. The data corpus used for their study is quite small and malware detection rate achieved is relatively low. In Maiorca {\textit{et al.}}'s work \cite{MaiorcaGC12}, they developed a tool termed PDFMS based on static analysis of PDF data structure. It comprises a data retrieval module, a feature extractor module and the classifier itself. It first identifies important features (keywords) using a K-means clustering technique, then Bayesian, SVM, J48 and Random Forests based algorithms are studied with Random Forest method performing the best. An alternative solution based on static analysis of PDF metadata and structural featuresis is PDFrate proposed by \cite{SmutzS12C}. In their study, Random Forest classifier yields the best detection rates as compared to other methods evaluated. In 2013, Srndic and Laskov published their new research results on PDF malware detection using decision tree and SVM algorithms \cite{SrndicL13C} in which an off-the-shelf PDF parser Poppler \cite{Poppler} is used to extract features. As compared to their previous work which focuses on detection of JavaScript content embedded in PDF files \cite{LaskovS11C}, the newly proposed approaches do not suffer from this limitation but have difficulty to handle sudden changes of attacks. More recent study on this topic can be seen in \cite{CuanDDV18T} using an SVM based model, but the associated low number of features and small dataset can affect their model generalization.

In this paper, we propose a novel approach based on an MLP neural network model, termed MLP$_{df}$, for the detection of PDF based malware. More specifically, the MLP model uses a backpropagation (BP) algorithm with stochastic gradient decent (SGD) search for model update. The model is trained and evaluated with structural properties, metadata and content information extracted from two datasets which contain around $105000$ real-world malicious and benign PDF documents. Our evaluation has demonstrated excellent behaviour of our approach, which greatly outperforms selected well known commercial anti-virus (AV) scanners, achieving a much higher true positive rate (TPR) while maintaining a very low false positive rate (FPR). The remainder of this paper is organized as follows: In the following section, the proposed MLP$_{df}$ method is introduced, including description on the MLP model and feature engineering. Then, Section \ref{results} contains evaluation results illustrating the performance of the MLP$_{df}$ method. Finally, Section \ref{concl} contains our conclusions and future work suggestions.

\section{Proposed MLP$_{df}$ Approach}\label{MLPdf}
\subsection{MLP Model}\label{MLP}
An MLP is a class of feedforward ANN with an input layer, an output layer, and one or more hidden layers between them. Each node in one layer fully connects to every node in the following layer. Except for nodes in the input layer, each node is a neuron (or processing element) with a nonlinear activation function associated with a scalar weight which is adjusted during training. An MLP is a supervised learning algorithm that learns to identify data patterns for classification or regression. In order to carry out the learning process, one needs to extract a digital representation $\textbf{x}$ of a given object or event that needs to be fed into the MLP. The learning task becomes to find a multidimensional function $\Phi(\cdot)$ between input $\textbf{x}$ and target $\textbf{y}$

\begin{equation}\label{eq:ml}
\textbf{y} \cong \Phi(\textbf{x})
\end{equation}
where $\textbf{x} \in \mathbb{R}^{N}$, a real-valued input feature vector $\textbf{x} = [x_{1}, \cdots, x_{N}]^{T}$ in an $N$ dimensional feature space, with $(\cdot)^T$ denoting the transpose operation. Similarly, $\textbf{y} \in \mathbb{R}^{M}$, a real-valued target classification vector $\textbf{y} = [y_{1}, \cdots, y_{M}]^{T}$ in an $M$ dimensional classification space. In other words, an MLP learning process is to find $\Phi(\cdot)$ which maps the data from feature space to classification output space. For a binary classification problem like PDF attacks detection, $M = 1$ and $y$ is a scalar (a vector with length $1$). 

\begin{figure}[htbp]
\centering
\begin{tikzpicture}
[   cnode/.style={draw=black!30,fill=#1,minimum width=3mm,circle},
]
   \node at (0,-2.8) {\includegraphics[scale=0.15]{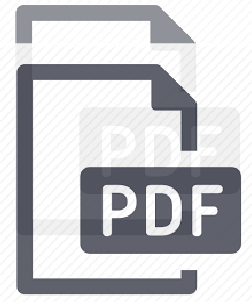}};
   \node[cnode=red,label=0:$\hat y \in {(0,1)}$] (s) at (7,-3) {};
   \node at (1.5,-3.8) {$\vdots$};
   \node at (3.5,-3.8) {$\vdots$};
   \node at (5.5,-3.8) {$\vdots$};
   \foreach \x in {1,...,4}
   {   \pgfmathparse{\x<4 ? \x : "N"}
       \node[cnode=blue,label=180:$x_{\pgfmathresult}$] (x-\x) at (1.5,{-\x-div(\x,4)}) {};
       \pgfmathparse{\x<4 ? \x : "J"}
       \node[cnode=gray,label=90:$u^{1}_{\pgfmathresult}$] (p-\x) at (3.5,{-\x-div(\x,4)}) {};
       \pgfmathparse{\x<4 ? \x : "K"}
       \node[cnode=gray,label=90:$u^{2}_{\pgfmathresult}$] (z-\x) at (5.5,{-\x-div(\x,4)}) {};
       \draw [-{latex[black!50]}] (z-\x) -- node[above,sloped,pos=0.3]{} (s);
   }
   \foreach \x in {1,...,4}
   {   \foreach \y in {1,...,4}
       {   \draw [-{latex[black!50]}](x-\x) -> (p-\y);
           \draw [-{latex[black!50]}](p-\x) -> (z-\y);
       }
   }
\end{tikzpicture}
\caption{Overview of MLP Architecture}
\label{fig-mlp}
\end{figure}
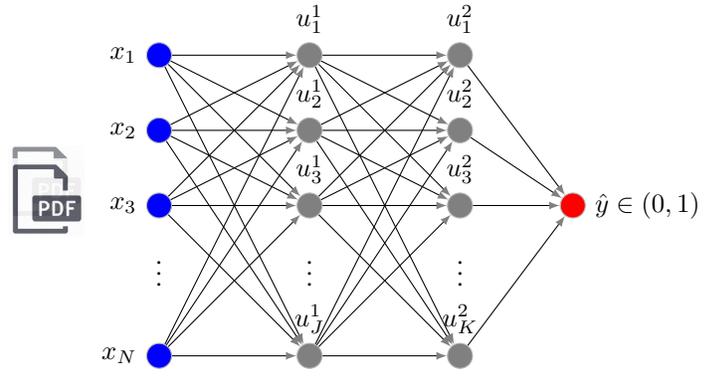

Fig. \ref{fig-mlp} illustrates the overview of an MLP network for a binary classification problem, where $\textbf{x}$ and $\hat y$ denote input feature vector and trained binary output, respectively, with two hidden layers. The number of layers used in the figure is for illustration purpose only, there could be more or less hidden layers depending on applications. Without loss of generality, we denote $\hat y^{i}_{k}$ as the output of $u^{i}_{k}$, the neuron unit $k$ in layer $i$ where $i = 1,\cdots, L$ and $k = 1,\cdots, K$. Note $i$ starts with $1$ as conventionally the input layer is layer $0$, and neuron units start with layer $1$ (the first hidden layer). The functionality of a neuron unit can be depicted below.

\begin{figure}[htbp]
\centering
\begin{tikzpicture}
[   cnode/.style={draw=black!30,fill=#1,minimum width=5mm,circle},
]
   \node[label=90:$\hat y^{i-1}_{j-1}w_{j_{i-1},k_{i}}$] (t-1) at (1,{-1-div(2,4)}) {};
   \node[label=180:$\hat y^{i-1}_{j}w_{j_{i-1},k_{i}}$] (t-1) at (1.5,{-2-div(2,4)}) {};
   \node[label=270:$\hat y^{i-1}_{j+1}w_{j_{i-1},k_{i}}$] (t-1) at (1,{-3-div(2,4)}) {};
   
   \node[cnode=white,label=90:$u^{i}_k$] (p-2) at (3.5,{-2-div(2,4)}) {$f^{i}(\cdot)$};
   \node[label=0:$\hat y^{i}_{k}$] (t-2) at (5.5,{-2-div(2,4)}) {};

   \draw [-{latex[black!50]}](x-1) -> (p-2);
   \draw [-{latex[black!50]}](x-2) -> (p-2);
   \draw [-{latex[black!50]}](x-3) -> (p-2);
   
   \draw [-{latex[black!50]}](p-2) -> (z-1);
   \draw [-{latex[black!50]}](p-2) -> (z-2);
   \draw [-{latex[black!50]}](p-2) -> (z-3);
   
\end{tikzpicture}
\caption{Functionality of a neuron in an MLP}
\label{fig-acf}
\end{figure}
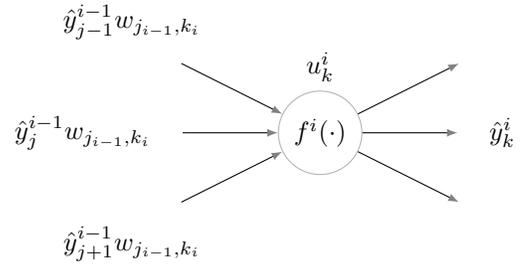
As Fig. \ref{fig-acf} shows, each neuron unit $u$ is a perceptron which realizes a transformation of the signal via application of its activation function $f(.)$ to its argument. Different activation functions can be used in different layers or even for different neurons in a layer. Rectified linear unit (ReLU), Sigmoid and Softmax are common choices for nonlinear activation functions \cite{Mitchell97}. The transformation resulting in the activation output as input for neurons of next layer can be formed as 

\begin{equation}\label{eq:acf}
\hat y^{i}_{k} = f^{i}(\sum_{j=1} \hat y^{i-1}_{j}w_{j_{i-1},k_{i}} + b^{i}_k)
\end{equation}
where $f^{i}(\cdot)$ is a nonlinear activation function, $w_{j_{i-1},k_{i}}$ the adjustable weight applied to the signal between unit $j$ in layer $i-1$ and unit $k$ in layer $i$, and $b^{i}_k$ a bias term. At input layer, $\hat y^{0}_{j} = x_j$.

\begin{figure}[htbp]
\centering

\begin{tikzpicture}[auto, node distance=2cm,>=latex']
   \node [input, name=input] {};
   \node [block, right of=input] (system) {MLP};
   \node [output, right of=system] (output) {};
   \node [block, below of=system] (sgd) {BP with SGD};

   \draw [draw,->] (input) -- node {$\textbf{x}$} (system);

   \node [sum, right of=system, pin={[pinstyle]above:$y$},
           node distance=2.5cm] (sigma) {$e(\hat y, y)$};
   \draw [->] (system) -- node {$\hat y$} (sigma);
   \draw [->] (sigma)  |- node {$e$}(sgd);

   \draw [->] (sgd) -- node {$\{w_{j,k}\}$} (system);
\end{tikzpicture}
\caption{BP based MLP weights update via SGD search}
\label{fig-bp}
\end{figure}
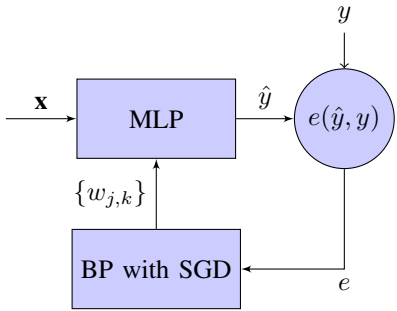

The goal of the MLP learning process is to find a set of optimal weights $\{w_{j,k}\}$ over a training dataset in order to bring the outputs as close as possible to the target values. The BP based MLP employs SGD search through the space of possible weight values to minimize the error signal between trained outputs and target values, as depicted in Fig. \ref{fig-bp}, where $e(\hat y, y)$ is the cost function for error measurement and the SGD based weight update rule becomes
\begin{equation}\label{eq:Delta}
\Delta w_{j_{i-1},k_{i}} =  - \eta \frac{\partial e(\hat y, y)}{\partial w_{j_{i-1},k_{i}}}
\end{equation}
where $\eta$ is the learning rate (step size), and $\frac{\partial e(\hat y, y)}{\partial w_{j_{i-1}},k_{i}}$ is the partial derivative of $e(\hat y, y)$ with respect to $w_{j_{i-1},k_{i}}$. The cost function $e(\hat y, y)$ can be defined in different ways as long as it is  differentiable so SGD can be applied. For binary probabilistic classification as the case of PDF malware detection, the cross-entropy of the network is a proven effective cost function for an MLP with a Sigmoid output, as defined below \cite{Mitchell97}
\begin{equation}\label{eq:cost}
e(\hat y, y) = -y \log{\hat y} - (1-y) \log(1 - \hat y)
\end{equation}
The MLP learning process starts with initial random weights assigned to $\{w_{j,k}\}$ and jointly apply slight modification $\{\Delta w_{j_{i-1},k_{i}}\}$ to the weights based on the computed gradient (as in (\ref{eq:Delta})) of loss $e(\hat y, y)$ after processing one data sample or a batch of samples at a time, over the whole training dataset. A complete training cycle is termed as an \emph{epoch} and many \emph{epoch}s are needed in order to accomplish an ANN training task. The learning process stops when it reaches the end of predefined \emph{epoch}s or $e(\hat y, y)$ is below a certain threshold. The full derivation of weight update rule in (\ref{eq:Delta}) is not discussed here, interested readers can refer to \cite{Mitchell97,GoodfellowBC16} for more details. 

\subsection{Feature Engineering}\label{FEng}
For classification learning problems such as PDF malware detection discussed herein, it requires the training datasets to be labeled as malware or benign first. Furthermore, as discussed in Section \ref{MLP}, data features rather than the raw data are used as input signal for the MLP model. One should not expect an MLP model to be able to learn from completely arbitrary data. The process of feature extraction is called feature engineering, an important part of data preprocessing for ML. It normally requires domain knowledge to identify related features from raw data. There is no exception when dealing with the detection of PDF attacks. Good features can simplify the relationship between input data and target values of the problem that an ML algorithm is being used to solve. One could argue that modern DL algorithms like convolutional neural network (CNN) are able to automatically extract features and learn from raw data without special knowledge of the problem. Such methods do have improved the capability of learning models to find hidden features, but it is unrealistic to believe that they can discover any structure in any type of data. Furthermore, it is not guranteed that CNN-like methods can perform better than manual feature engineering based approaches such as MLP \cite{ShaheenVA16C}.

It is essential to have good quality features as input, as incorrect choices of features can lead to an ML classifier to perform less well or completely fail. In this demonstration, a group of selected $48$ high quality features are extracted from both training and testing datasets shown in Table \ref{tbl:feat}. In-house tools and off-the-shelf PDF parsers are used for the feature extraction. These features range from PDF structure information, object characteristics, metadata information to content statistical properties. Apart from features like JavaScript with/without obfuscation, number of objects, page count, stream filtering, other important structural and content information can deliver strong indication for the presence of malicious content in a PDF file. For example, a feature can be defined for possible embedded files in a PDF file and another feature for entropy of some content. All such features bear a strong discriminative power to differentiate malicious documents from benign ones. For commercial reasons, the full list of features and how they are extracted are not discussed in this paper.

As part of data preprocessing, all extracted features need to be in the form of numerical values which an ML network can only digest. Further data processing is also required before feeding the features into an ML model, which is discussed in the following section.

\begin{table}[hhtb]
\caption{Datasets and feature information}
\begin{center}
\begin{tabular}{|c|c|c|c|c|}
\hline
\multicolumn{2}{|c|}{\textbf{Training: $90000$}} & \multicolumn{2}{|c|}{\textbf{Testing: $15047$}}&\textbf{Features: $48$}\Tstrut\Bstrut\\
\hline
\textit{Benign}& \textit{Malicious}& \textit{Benign}& \textit{Malicious}& Structure, object properties \Tstrut\Bstrut \\
\cline{1-4} 
\Tstrut\Bstrut
\textit{$78684$}& \textit{$11316$}& \textit{$13101$}& $1946$ & metadata, content stats, etc. \Tstrut\Bstrut\\
\hline
\end{tabular}
\label{tbl:feat}
\end{center}
\end{table}

\section{Evaluation Results}\label{results}
In this demonstration, the training and testing tasks are carried out with datasets of around $105,000$ real-world benign and malicious PDF documents, as shown in Table \ref{tbl:feat}. The majority of the malicious PDF files are from Sophos malicious PDF document collections over a few months time period up to March 2018, while the collection of benign files has a longer timespan. In general, the benign to malware ratio in a real-world situation is much higher than $1$, therefore we adopt a reasonable high ratio of around $6$ for our training and testing datasets to reflect the real situation. 

\subsection{Data Preprocessing}\label{prep}
In addition to the feature extraction discussed in Section \ref{FEng}, there involves further data preprocessing before feeding the numerical features into a network. The common process includes \emph{Normalization} and \emph{Regularization}, as discussed below.

\begin{itemize}
\item \emph{Normalization} - The purpose of applying normalization to features is to avoid large gradient updates which might prevent the learning algorithm to converge. Typically the feature vectors will be normalized independently so that each feature vector will be scaled to have a standard normal (or Gaussian) distribution with $\mu=0$ and $\sigma=1$ where $\mu$ and $\sigma$ are the mean and standard deviation of a normalized feature vector, respectively. This will make it easier for an ML algorithm to learn. In our work, all the feature vectors have been normalized this way before feeding them into the model;
\item \emph{Regularization} - One of the most challenging problems an ML algorithm faces is how to tackle overfitting. In other words, a trained algorithm must perform well on new data, not just those trained data. There exist various ways to mitigate this issue \cite{Mitchell97,GoodfellowBC16}, of which \emph{Dropout} and \emph{Batch normalization} are widely used. In our evaluation, a dropout rate of $0.15$ is used, which means $15\%$ of each hidden layer outputs are zeroed out before feeding into next layer. It's worth noting that dropout should only be applied during training process, not testing or production stage. To make MLP$_{df}$ model learn better during training and generalize well on new data, a batch size of $64$ is used and batch normalization is applied. This is similar to feature vector normalization for input layer, the goal is to re-scale each layer's input to have zero mean and unit variance. In addition, a validation dataset comprising $20\%$ of the training dataset is used to help detect overfitting and perform model selection during training process. 
\end{itemize}

\begin{figure}[htbp]
\centerline{\includegraphics[scale=0.6]{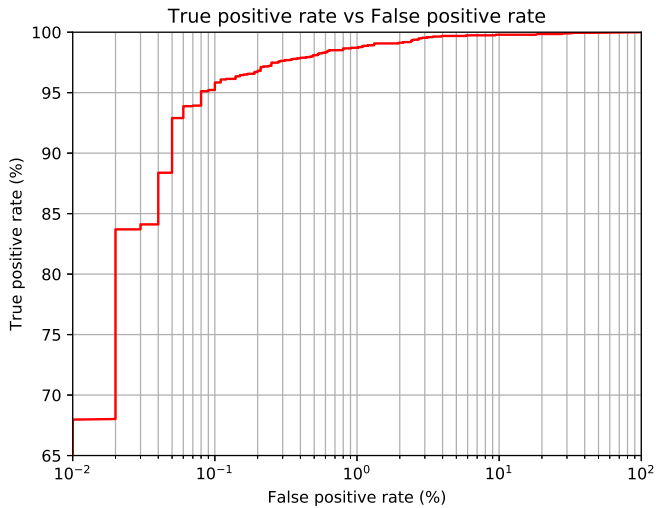}}
\caption{Testing results: true positive rate vs false positive rate.}
\label{fig:tp_fp}
\end{figure}
\subsection{Testing Results}\label{testing}
The testing results presented in this section are based on an optimal trained model which has two hidden layers (similar to the one shown in Fig. \ref{fig-mlp}). Each hidden layer in the model has $72$ neurons. The input layer has $48$ nodes corresponding to the number of features used, and the output layer has a single Sigmoid (binary) probability output with values in the range of $(0, 1)$. The model is trained with $5000$ \emph{epoch}s. Fig. \ref{fig:tp_fp} shows how a TPR changes over a varying FPR. As it implies, when an FPR is close to $0.01\%$ (or $1e^{-4}$), the related TPR is below $70\%$. A slightly higher FPR can result in a significantly improved TPR. For example, a TPR of $95\%$ can be achieved with an FPR near $0.1\%$ (or $1e^{-3}$). 
\begin{figure}[htbp]
\centerline{\includegraphics[scale=0.6]{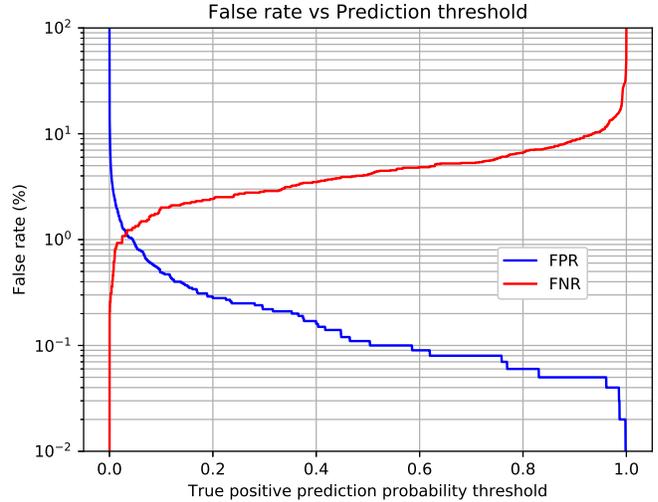}}
\caption{Testing results: false rate over varying probability threshold.}
\label{fig:fp_fn}
\end{figure}
The correlation of TPR and FPR is related to the model output which is a prediction probability for a PDF file being benign or malicious. A modified prediction probability threshold can lead to a different FPR and a changed false negative rate (FNR), as depicted in Fig. \ref{fig:fp_fn}. 

\begin{figure}[htbp]
\centerline{\includegraphics[scale=0.6]{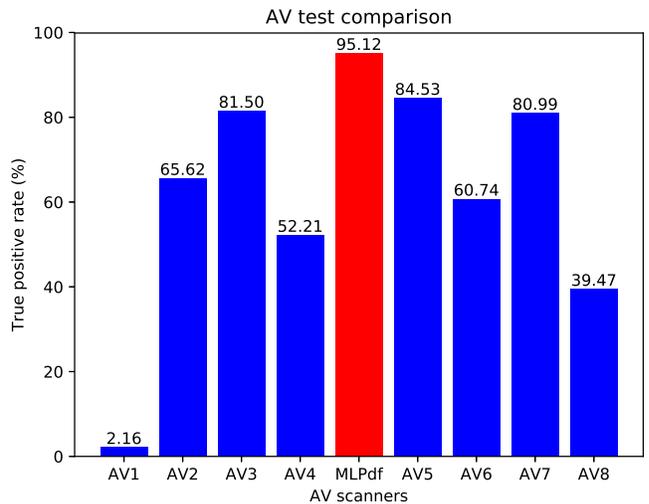}}
\caption{Testing results between MLP$_{df}$ and major AV scanners.}
\label{fig:av}
\end{figure}
As the figure shows, when a low true positive prediction threshold is selected, a high FPR and a low FNR are expected. On the other hand, as the threshold increases, the FPR decreases with an increasing FNR. An aggressive model can demonstrate remarkable detection accuracy (a low FNR), but at the cost of a much higher FPR. It is important to have a good balance with high detection rates and minimal false positives. 

The probability of $0.62$ is used as the true positive prediction threshold for the MLP$_{df}$ model, which results in a very low FPR of $0.08\%$ and an excellent $95.12\%$ TPR (or $4.88\%$ FNR) based on $13101$ benign and $1946$ malicious PDF documents from the testing dataset listed in Table \ref{tbl:feat}. The MLP$_{df}$ true positive prediction result along with the results from eight major commercial AV scanners are shown in Fig. \ref{fig:av} where AV-$1$ to AV-$8$ denote the corresponding AV scanners. It clearly shows that the MLP$_{df}$ approach (red bar) significantly outperforms all commercial scanners with a big margin. The best commercial scanner only has a TPR of $84.53\%$ while the MLP$_{df}$ achieves over $95\%$ TPR.

As mentioned above, the MLP$_{df}$ approach maintains a low FPR of $0.08\%$ for the $13101$ benign testing files based on the true positive prediction probability of $0.62$. The evaluated eight commercial scanners perform well with zero or very low FPR as well. Part of the reason could be that the majority of our benign files are collected from Sophos benign PDF documents datasets which have a relatively longer timespan, and most of the files might be already known to other commercial scanners as well. 

\section{Conclusion}\label{concl}
In this paper, we have proposed a novel approach based on a multilayer perceptron neural network model, termed MLP$_{df}$, for the detection of PDF based malware. More specifically, the MLP model uses a backpropagation algorithm with stochastic gradient decent search for the model update. The MLP$_{df}$ model has an input layer of $48$ nodes, two hidden layers with $72$ neurons in each layer and a single Sigmoid output layer. A group of selected $48$ high quality features are extracted from two real-world datasets which comprise around $105000$ benign and malicious PDF documents. Testing results indicate that the proposed MLP$_{df}$ approach exhibits excellent performance which significantly outperforms all evaluated eight well known commercial anti-virus scanners with a very high TPR of $95.12\%$ achieved while maintaining a very low FPR of $0.08\%$. It is worth noting that the commercial scanners perform well on the benign testing files as well with zero or very low FPR. As part of the future work, it will be interesting to compare how MLP$_{df}$ and other commercial scanners perform with a larger data corpus, particularly adding more recent benign PDF documents to the dataset.


\end{document}